# Case Study on Using Colours in Constructing Emotions by Interactive Digital Narratives

Kuldar Taveter[1] and Eliise Marie Taveter[2]

[1] Institute of Computer Science, University of Tartu, Estonia
[2] University of Nottingham, United Kingdom
kuldar.taveter@ut.ee

**Abstract.** This article addresses the possibility of supporting the construction of emotions in the participants of Interactive Digital Narratives (IDN) by means of colours. The article uses goal models for expressing protostories. The core of the article consists of the case study where two colour synesthetes were asked to choose colours for eight emotions. Thereafter the same synesthetes were asked to choose colours for the emotions that support the attainment of goals in the Cinderella narrative, which serves as an example protostory. The article also discusses the perspectives for applying the method proposed by us to using colours in constructing emotions by IDNs.

## 1 Introduction

This article is concerned with constructing emotions by Interactive Digital Narratives (IDN). IDN has been defined as an "expressive narrative form in digital media implemented as a computational system containing potential narratives and experienced through a participatory process that results in products representing instantiated narratives" [1]. IDN consists of a protostory and narrative design [1]. *Protostory* determines the IDN's potential set of instantiated narratives [2]. In this sense protostory is ontologically a *type* that carries a principle of identity for its instances and whose every instance maintains its identity in every circumstance considered by the model [3]. Differently from creating instances of object types in programming languages, where only the data content of object instances varies, also the behaviours of the protostory instances are different, depending on the actions by the IDN participants at runtime. It is emphasized in [2] that a protostory is "more than just a computer program, as the term encompasses not only the concrete programming code and interactive interfaces, but also the artistic intent that enables a participatory process of instantiation that results in the realization of potential narratives". *Narrative design* is the structure within a protostory that contains and enables a flexible presentation of a narrative [2]. This includes the segmentation of the protostory into smaller self-contained fragments of the narrative, sequencing of the fragments and creating the connections between them, including the points where participants can interfere [2].

Emotions play an important role in IDNs. Furthermore, IDNs can even be designed to construct in participants emotions of a certain kind. According to Rieser [4], "the responsive nature of such systems opens up a potential new craft for the writer, where the encoding of mood, emotion and their syntax takes precedence over plot and traditional forms of narrative technique." For example, in the IDN "Muse" described in [4], a software agent talks poetry to which the audience responds in their own words. The animated "Muse" controlled through a neural network recognises emotional nuances in the phrases used by the audience and replies in turn with emotional expressions. Another example of emotion-intense IDN is Aspergion described in [5], which is a multiplayer online role play game that promotes respect for people with Asperger's syndrome.

This article approaches encoding of emotions in IDNs from the perspective of the theory of constructed emotion. According to the theory of constructed emotion [6], a human brain invisibly constructs everything one experiences, including emotions. Emotions are constructed in the brain, in concordance with the goals aimed to be achieved. For example, if the goal is romantic love, the emotions "Passionate", "Longing" and "Lustful" might be constructed which make this goal more attainable. Differently, if the goal to be attained is tough love or brotherly love, respective instances of the emotions "Disciplined" and "Bonded" might be constructed [7]. The brain may also combine existing emotions to construct an instance of a new emotion. For example, the Japanese emotional concept of "mono no aware" (物の哀れ) stands for the despair felt at the impermanence and transience of life, especially in its most satisfying moments when our love for others, and their need for us, feels so unexpectedly overwhelming – and life so very fragile and temporary – that we become very sad [8]. This is a good example of a complex emotion that is constructed from other emotions such as feeling love, sad and overwhelmed.

At the elementary level, emotions are constructed from basic affects, which are *not* emotions but much simpler *feelings* with two features: valence and arousal. *Valence* characterises how pleasant or unpleasant one feels, while the feature of *arousal* is how calm or excited one feels [9]. A person's momentary core affective state can be psychologically represented as the affective circumplex. The horizontal dimension, hedonic valence, ranges from pleasant states at one end to unpleasant states at the other. The vertical dimension, arousal, ranges from high activity and attention at one end to low activity and sleepiness at the other [10]. The immediacy of affects, and the subsequent bodily responses make them easier to detect by psychological tests or automatically using physiological measures; such as detecting excitement with a heart rate measure. Based on the valence and arousal of its constituent core affects, an emotion can also be characterised by its valence and arousal.

Emotions have been shown to be associated for a person with different colours [11-18]. This has been utilized also in IDNs. For example, the Hybrid Book [19] is an e-book where the text of a story is annotated with emotions that one or another fragment of the story should create in the listeners. When the book is being read by someone, based on the annotations, specific colours can be associated with the current "mood" or emotion of the story that is being told: e.g., the blue colour can be associated with "Sad". The Hybrid Book is connected to the Philips hue lights that have the functionality of changing the colour of lights based on the emotions prevailing in different fragments of the story. The work proposed in [19] is based on the six basic emotions proposed by Ekman [20]: “Angry”, “Afraid”, “Disgusted”, “Sad”, “Surprised”, and “Happy”.

Likewise, the article [21] has analysed by experimentation the effect of colours in movie-based IDNs. An experiment was conducted where viewers of an interactive movie could choose along with the next movie fragment to be watched the colour calibration for the next fragment. This research reported in [21] demonstrates that the choice of colour calibration actually decides the outcome of a movie-based IDN. The usage of colours in movies has already become more important than the factors such as simple screen composition, characters of main actors, and environment where the movie has been shot and has therefore obtained an important role also in movie-based IDNs. Considering this, the usage of colour in movie-based IDNs is an important topic that should be studied and further developed in movie making.

This article is structured as follows. In Section 2, we describe the motivation for performing the research work reported in the article, state the research goals, and provide an overview of the research methods employed by us. In Section 3, we describe the results of the case study in terms of the colours chosen by the participants, the colours chosen or modified by them in the context of the example narrative, which can be viewed as a protostory, and map the colours chosen by the participants to the circumplex model of emotion. Finally, in Section 4 we draw conclusions and discuss the perspectives of applying the method of using colours in constructing emotions by IDNs.

## 2 Methodology

**Motivation.** Synaesthesia is a condition in which one type of stimulation evokes the sensation of another, as when the hearing of a sound leads to the perception of colours [23]. Numerous articles (e.g., [23, 29-33]) have reported strong emotional responses to colours by people with synaesthesia. Despite of that, the experience of emotions associated with synesthetic events is a commonly known but poorly studied feature of grapheme-colour synaesthesia [33]. To fill this gap, a new series of studies on “affective synaesthesia” would be needed [34]. As such studies are not yet available, our justification for conducting case studies with synesthetes for finding how colours are associated with emotions is based on indirect data about the associations between colours and emotions among creative people, as compared with others. The results of the tests reported in [35] clearly indicate that the synesthetes scored significantly higher than the control subjects on all four measures of creativity – Adjective Check List Creativity Scale [36], Barrow-Welsh Revised Art Scale [37], Obscure Figures Test [38], and Similes Test [39].

On the other hand, in the study reported in [40], high-creative and low-creative participants were asked to associate colours with emotions. Before that, the Remote Associates Test [41] was used to measure creative potential of the participants. The results of the study [40] clearly show that more creative participants are better at distinguishing associations between colours and emotions. The colours associated with four emotions by high-creative and low-creative participant are respectively depicted in Figures 1 and 2. The data for the figures originates in [40], where the colours associated with emotions are presented in terms of the ratings of mean similarity.

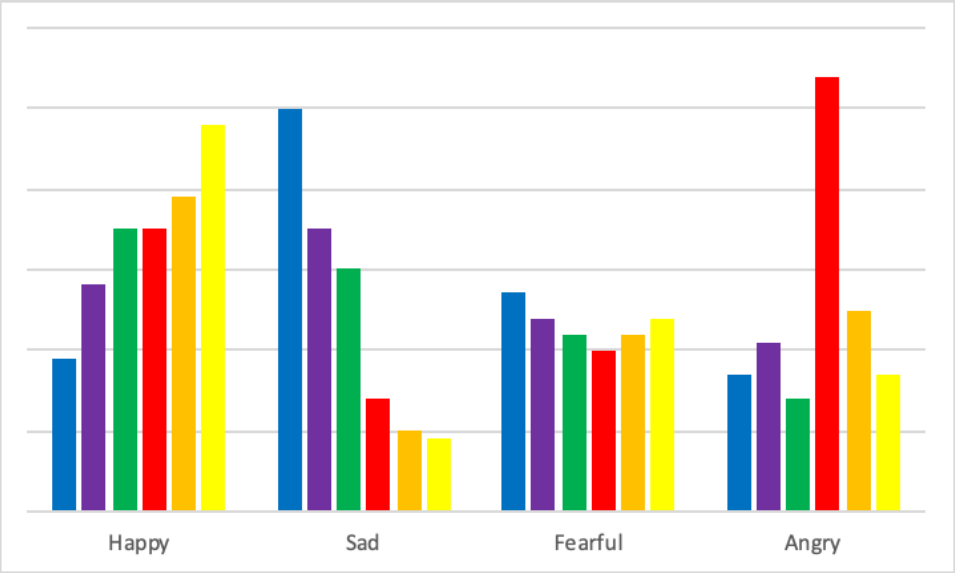

**Fig. 1.** The colours associated with emotions by high-creative participants. Source: By the author based on [40].

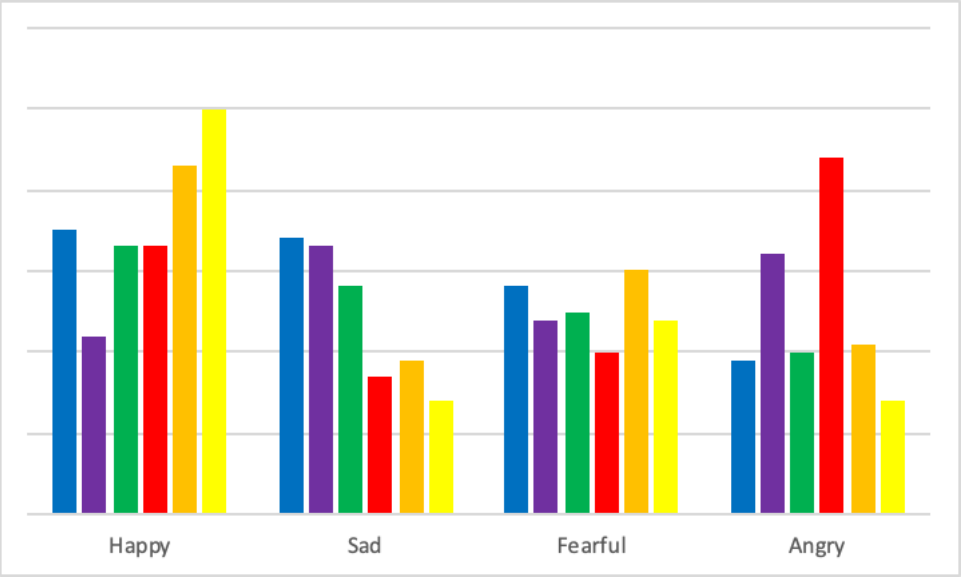

**Fig. 2.** The colours associated with emotions by low-creative participants. Source: By the author based on [40].

**Research goals.** Our main research goal was: *How are the colours that are used for representing different emotions associated with each other?* The sub-research goals were:

- What colours are used for representing which emotions?
- How stable are the colours used for representing one or another emotion?

**Research methods.** Broadly speaking, the research method employed by us is case study research [49]. We conducted a case study with two persons with colour synaesthesia [22-25] to whom we had access and who were willing to participate in the case study. The most common form (with a 64.4% prevalence among synesthetes [26]) of synaesthesia is grapheme–colour synaesthesia, in which achromatic letters or digits automatically trigger an idiosyncratic perceptual experience of colours (e.g., the letter 'j' is perceived as red) [27]. The second most prevalent form of synaesthesia (22.4% according to [26]) is time unit (e.g., Monday, January) colour synaesthesia [28]. Inducers of grapheme–colour synaesthesia are occasionally accompanied by the perceptions of shape, texture, and/or movement features, and even nonvisual perceptions such as smells and tastes, and particularly emotions [22]. One of the participants of the case study has grapheme–colour synaesthesia and the other participant has in addition time unit colour synaesthesia, which is accompanied by locating time units in space.

In the case study conducted by us, we first asked the two synesthetes to associate colours with emotion terms. We first asked the synesthetes to choose colours for the so-called six basic emotions [20] – "Happy", "Sad", "Fearful", "Angry", "Disgusted", and "Surprised" – and additionally, also for "Pleasant" and "Compassionate". It is important to note here that in compliance with the theory of constructed emotion [6], we consider basic emotions as core affects [9] and psychological primitives [10], out of which more complex emotions are constructed rather than as biological categories [7].

As the next step, we asked the synesthetes to choose the colours for emotions in the context of a narrative. For simplicity, we represented the narrative as a goal model. Goal models have been proposed by Sterling and Taveter in [43] and refined in [44]. The notation for representing goals models is shown in Figure 3. As the figure reflects, goal model can be considered as a container of three components: agents performing certain roles, functional goals, and quality goals. In goal models, agents performing certain roles are active entities that are required for achieving the goals set for the system, which in our case is IDN. Agent can be a character in the narrative, such as Cinderella, or the role required for enacting the IDN, such as Storyteller, as well as different participant roles that members of the audience can take. Functional goals or *do*-goals answer the question "what should be done or achieved?" and quality goals or *be*-goals answer the question "how the thing to be done or achieved should be?" or "what are its quality aspects?". Quality aspects also comprise the emotions that should be constructed to support the attainment of functional goals, which complies well with the theory of constructed emotion [6-7]. The skeleton of a goal model is a hierarchy of functional goals drawn as a tree. The hierarchical structure is to show that achieving the sub-goal represents a certain aspect of achieving its parent goal. The root of the tree sits at the top of the goal model and represents the overall goal of the system – the IDN. In goal models, quality goals and roles are attached to functional goals, as is indicated by Figure 3, whereby a quality goal or role attached to a functional goal recursively applies to all of the sub-functional goals of the given functional goal. In the goal model, the time runs from left to right and within each branch of the goal tree, from top to bottom.

The usage of goal models in creative domains of creating animations and serious games is described, for example, in the articles [45-47]. Since goal models are generic and flexible by representing *what* and *why* rather than *how* something should be done or achieved, they are appropriate means for representing protostories. In addition, goal model is an appropriate tool for representing emotion-intense IDNs because it explicitly models the goals to be achieved within an IDN and the emotions to be constructed to support the attainment of these goals. In our case study, we used as an IDN protostory the narrative that everyone knows – the story of Cinderella that was rendered as a goal model.

As the final step, we mapped the resulting colours for emotions to the circumplex model of emotion by Russell [42] shown in Figure 4. The model represents emotions in a two-dimensional space with the valence and arousal axes, stretching, respectively, from displeasure to pleasure and from low arousal to high arousal.

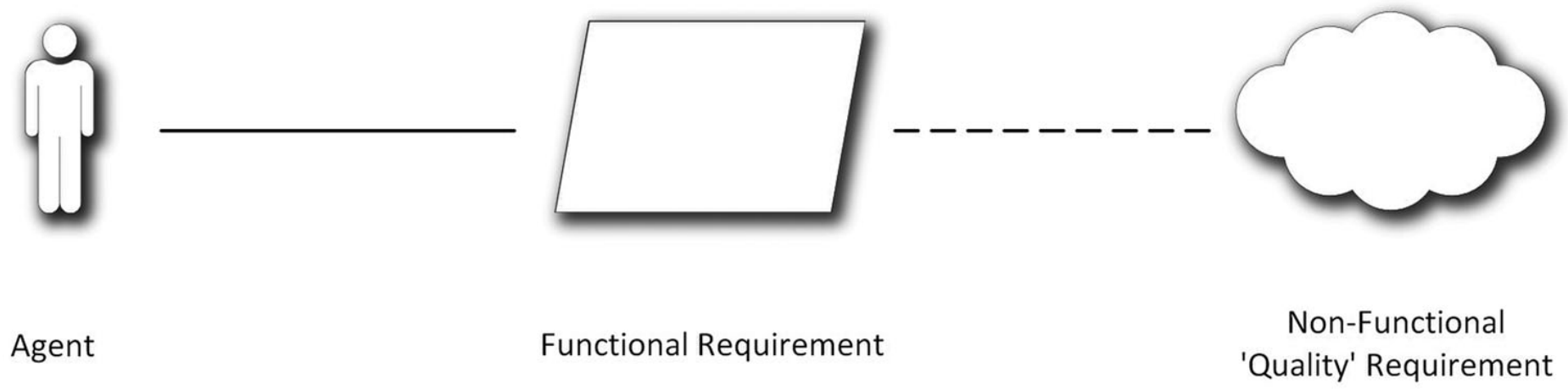

**Fig. 3.** The notation for goal modelling in the context of animations and serious games. Source: [45].

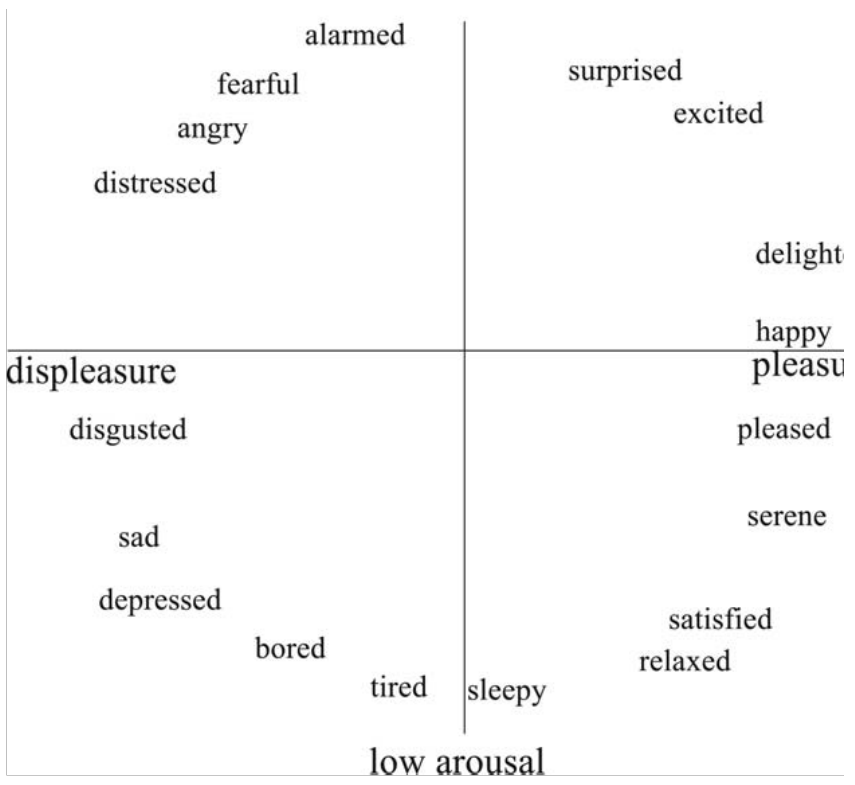

**Fig. 4.** Representation of emotions based on the circumplex model of emotion. Source: [11].

## 3 Results and Discussion

We conducted a case study with two persons with colour synaesthesia [22-25], who volunteered to participate in the case study. As was already described in Section 2, the case study was conducted in three stages. At the first stage, we asked the synesthetes to choose with the help of the colouring feature of the MS Word text editor colours for the emotions “Happy”, “Sad”, “Fearful”, “Angry”, “Disgusted”, “Surprised”, “Pleasant”, and “Compassionate”. The results of the first stage of the case study are shown in Table 1.

As Table 1 shows, the colours chosen by the Synesthete 1 and Synesthete 2 for a number of emotions are quite similar, while the colours chosen by them for the emotions “Happy”, “Surprised”, “Pleasant”, and “Compassionate” are noticeably different. Based on the article [12], it seems that the colours chosen for the emotions by the Synesthete 2 are relatively exceptional but this conclusion can be too far-reached because the article [12] is about native English-speakers but neither of the participating synesthetes is a native English-speaker.

**Table 1.** The colours chosen by the synesthetes for the emotions in the case study.

| Emotion/affect | Synesthete 1 | Synesthete 2 |
|---|---|---|
| Happy | 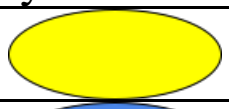 | 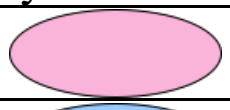 |
| Sad | 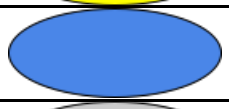 | 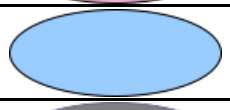 |
| Fearful | 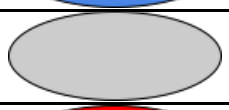 | 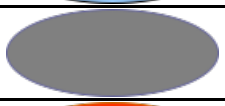 |
| Angry | 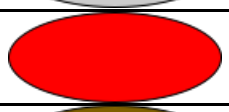 | 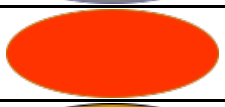 |
| Disgusted | 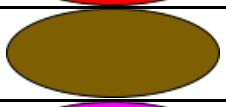 | 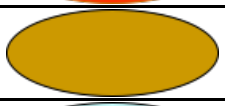 |
| Surprised | 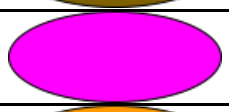 | 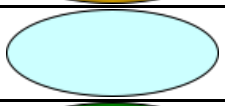 |
| Pleasant | 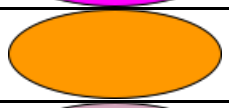 | 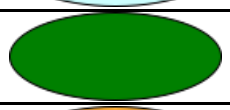 |
| Compassionate | 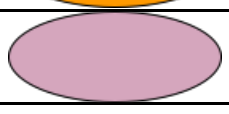 | 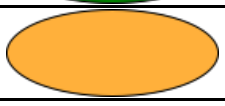 |

As we intended to study mapping emotions to colours in the context of IDNs rather than the mapping between emotions and colours *per se*, we next studied the colours assigned to the emotions by the synesthetes in the context of the Cinderella narrative – protostory. At stage 2, the participating two synesthetes were asked to choose with the help of the colouring feature of the MS Visio graphical editor colours for the quality goals capturing emotions that are associated with the functional goals representing the protostory. These quality goals represent how the audience of the story is supposed to feel when a given functional goal is being achieved.

The results of choosing by the Synesthete 1 colours for the emotions of the Cinderella protostory are shown in Figure 5. As Figure 5 reflects, in the context of the story, the colours for the emotions "Angry" and "Happy", which have been attached to a number of functional goals of the model, have changed as compared with the colours included by Table 1. According to the Synesthete 1, because of the context of the story, "Happy" has a subtle touch of "Angry" and the other way around. Accordingly, the Synesthete 1 represented "Happy" in a kind of light brown colour, which is a combination of yellow standing for "Happy" and red standing for "Angry". The resulting colour is somewhat close to "Pleasant" shown in Table 1. The colour used in Figure 5 for rendering the emotion "Angry" is a kind of brownish colour that is more tilted towards the red for "Angry" than the yellow for "Happy" shown in Table 1.

The Synesthete 1 described how she constructed colours for the emotions "Empathetic" attached to the functional goal "Lives an unhappy life" and "Sorry for her" attached to the functional goal "Loose one slipper". According to the Synesthete 1, the colour for "Sorry for her" is a combination of blue for "Sad" and red for "Angry", which results in a dark purple colour. This purple has different shades, depending on the functional goal to which it is attached but both shades, especially the one associated with the functional goal "Lose one slipper", are closer to the red for "Angry" than the blue for "Sad".

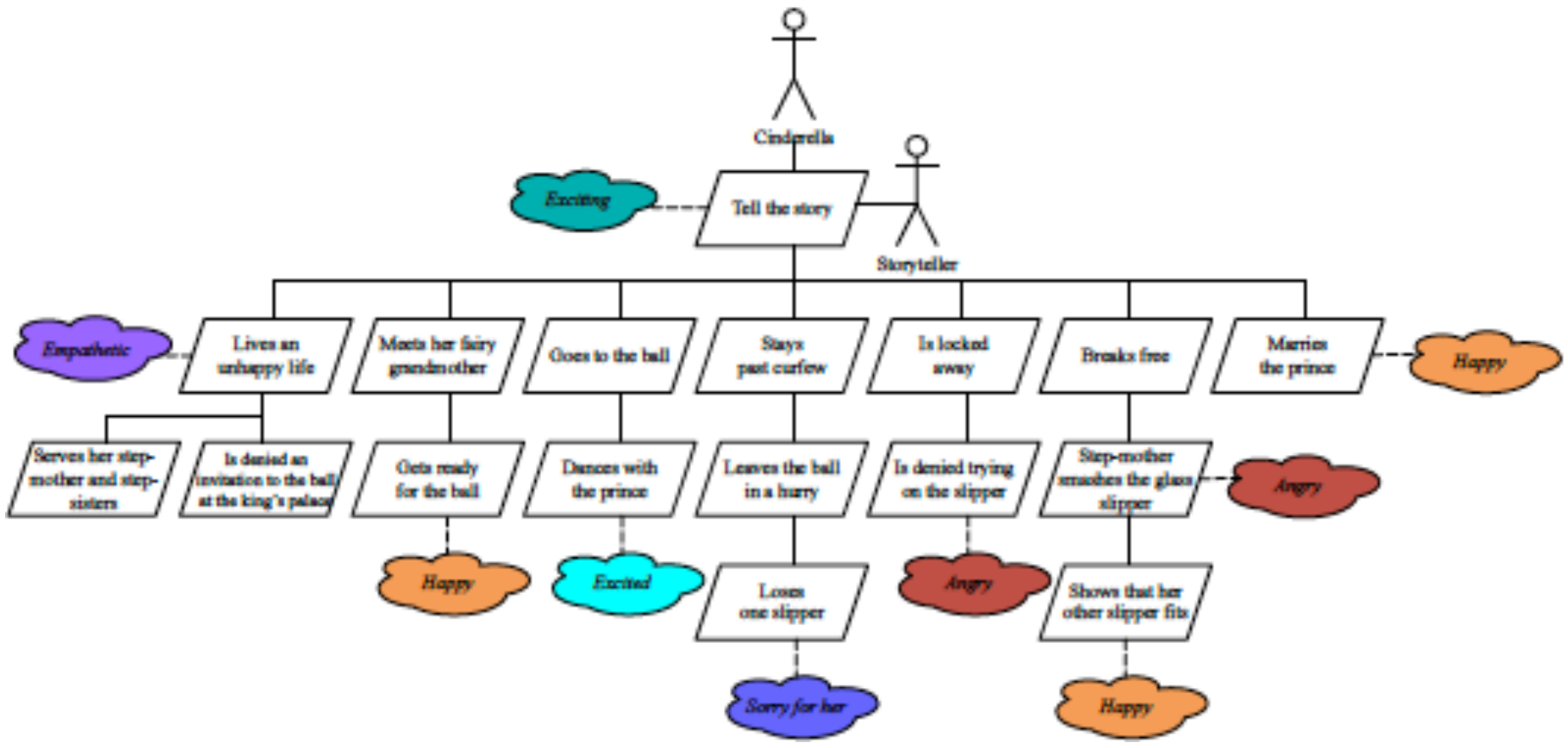

**Fig. 5.** The colours chosen for the Cinderella protostory by the Synesthete 1.

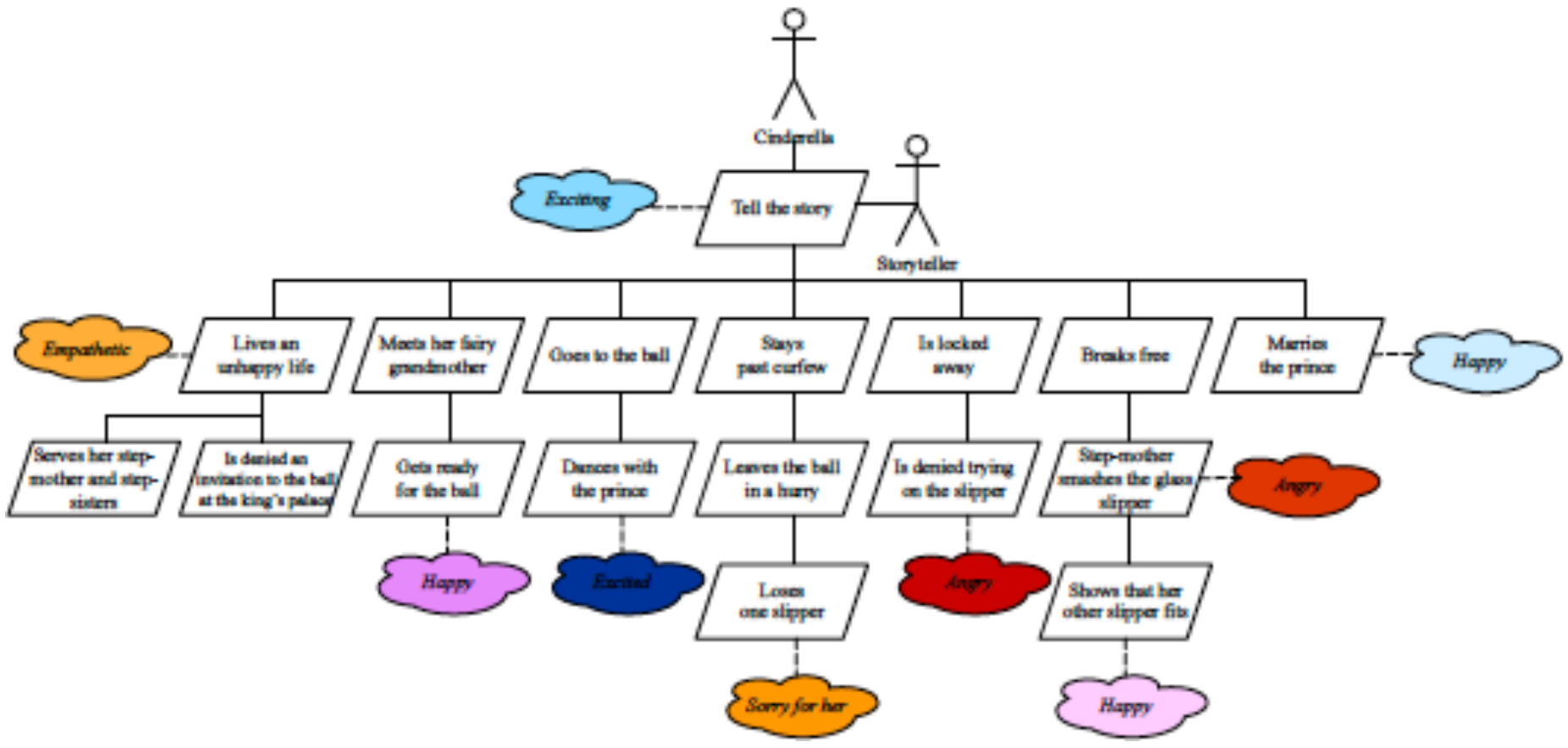

**Fig. 6.** The colours chosen for the Cinderella protostory by the Synesthete 2.

The Synesthete 1 provided a very interesting description of rendering the colours for the emotion "Exciting" connected to the functional goal "Tell the story" and the emotion "Excited" connected to the functional goal "Dance with the prince". According to the Synesthete 1, "the vibrant blue/green colour for the emotions 'Exciting' and 'Excited' is a combination of the yellow colour for the emotion 'Happy' and light blue colour for the emotion 'Calm', which represents for me being calm, cool, logically balanced and rational."

The results of choosing by the Synesthete 2 colours for the emotions of the Cinderella protostory are shown in Figure 6. The figure reflects that in the context of the narrative, the colours for the emotions "Angry" and "Happy", which have been attached to a number of functional goals of the model, mostly comply with the colours for the respective emotions shown for the Synesthete 2 in Table 1, with some slight differ-ences in shade. Similarly, the colour chosen by the Synesthete 2 for the emotion "Empathetic" attached to the functional goal "Lives an unhappy life" and for the emotion "Sorry for her" attached to the functional goal "Loses one slipper" comply with the colour for the emotion "Compassionate" shown in Table 1. However, there is one notable difference – the light blue colour chosen by the Synesthete 2 for the emotional goal "Happy" that is attached to the functional goal "Marries the prince". This colour for "Happy" seems to convey the emotion of "Surprised" (see Table 1) emphasized by the end of the protostory. A strong element of surprise is probably also the reason for a bit darker shade of blue that has been chosen by the Synesthete 2 for the emotion "Exciting" attached to the highest-level goal "Tell the story" of the goal model shown in Figure 6. The dark blue chosen by the Synesthete 2 for the emotion "Excited" attached to the functional goal "Dances with the prince" seems to be a kind of puzzle. An explanation can be that the colour of dark blue is a combination of cyan, which is close to the colour for "Surprised" chosen by the Synesthete 2, and magenta, which is close to the colour for "Happy" chosen by the Synesthete 2, as can be seen in Table 1.

For further study of the relationships between colours chosen for different emotions by both synesthetes, we mapped the colours for emotions chosen by them to the circumplex model of emotion by Russell [42]. The results of this mapping for the Synesthetes 1 and 2 can be seen in the respective Figures 7 and 8.

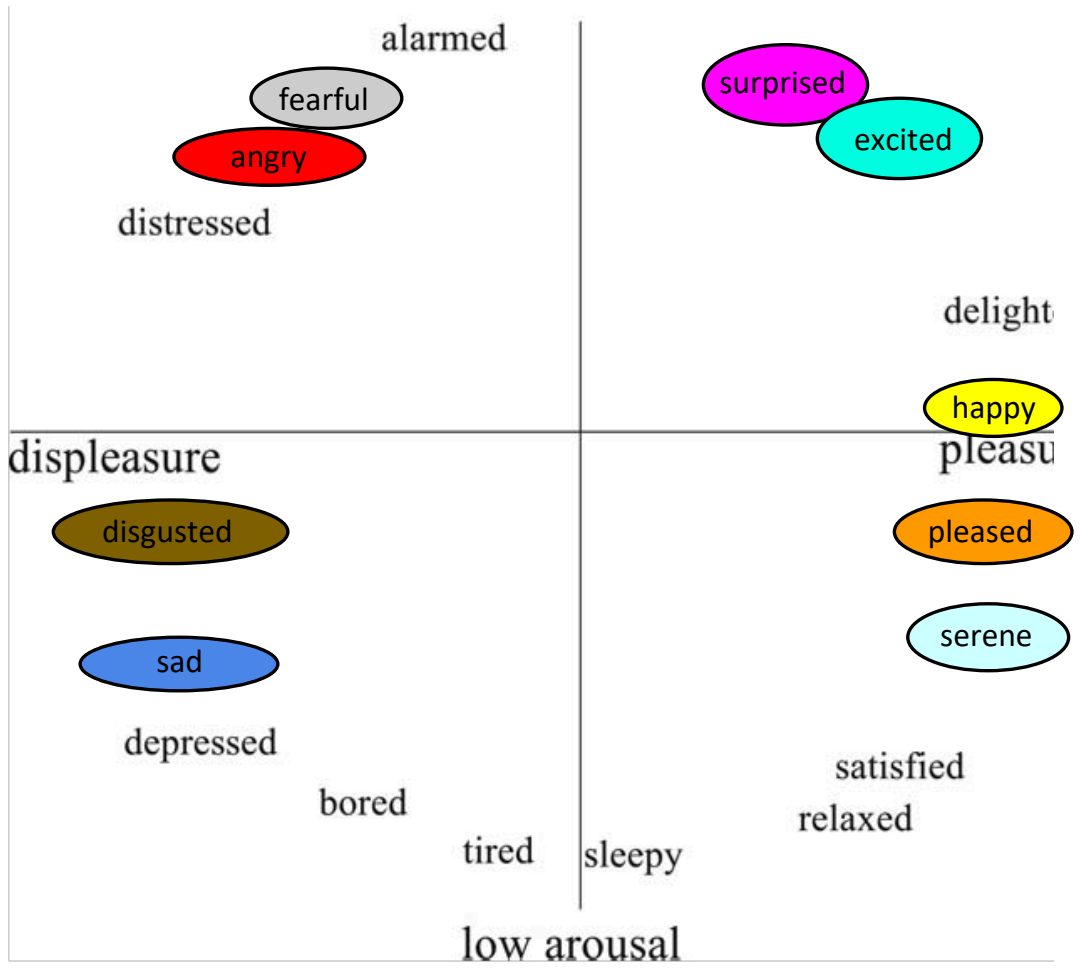

**Fig. 7.** Emotion colour mappings to the circumplex model for the Synesthete 1.

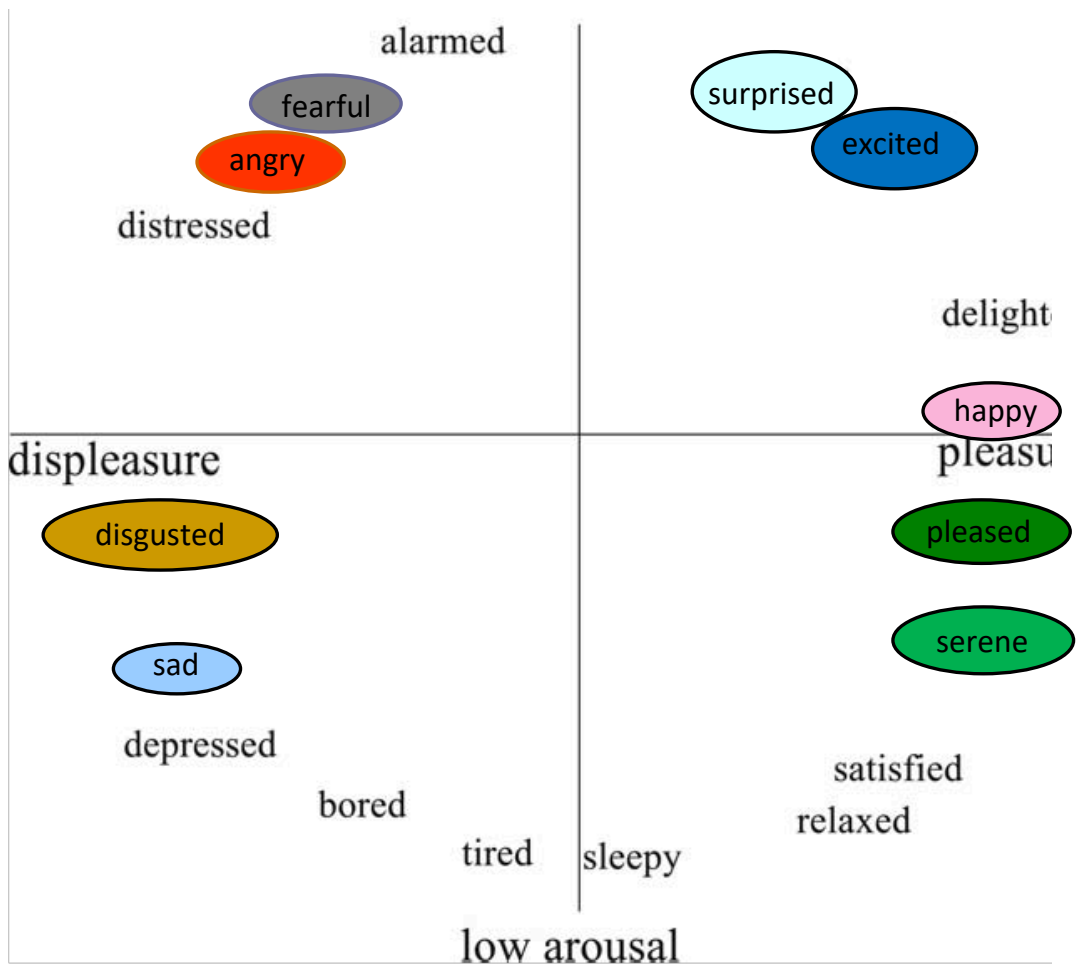

**Fig. 8.** Emotion colour mappings to the circumplex model for the Synesthete 2.

According to Figure 7, scales of hue can be observed along the horizontal displeasure-pleasure axis between the colours chosen by the Synesthete 1 for the emotion "Disgusted" on one end and "Pleased" and "Happy" on the other end, as well as between "Sad" and "Serene". The same kind of scale can be seen between the colours chosen by the Synesthete 1 for the emotions "Angry" and "Surprised".

Hue scales of the same kind cannot be detected for the Synesthete 2 but we can see a grouping of bluish colours around "Surprised" and "Excited" and a grouping of greenish colours around "Pleased" and "Serene".

The results of the case study clearly indicate that the assignment of colours to emotions is idiosyncratic. Also, as the synesthetes who participated in the case study, come from different cultural backgrounds, the assignment of colours to emotions is also culture-dependent. Both results were expected in the light of the theory of constructed emotion [6-7], which effectively claims that emotions are complex social constructs rather than something that people are born with.

The most interesting results of our case study are not concerned with the colours assigned by the synesthetes to the emotions as such but rather with the *order* or *system* how the synesthetes assigned colours to the emotions. Based on the results of the case study with the two synesthetes, we can claim that the colours that were chosen by either of the synesthetes for representing more complex emotions, such as "Excited", are systematically constructed from the colours chosen by the given synesthete for the so-called basic emotions, such as "Happy" and "Surprised". Moreover, the case study also demonstrates that the colours used by the synesthetes for representing the so-called basic emotions, such as "Happy", "Angry" and "Compassionate", remarkably depend on the context of the given protostory fragment.

While there are quite a few studies available on which colours are assigned by persons to different emotions [11-18], there are only a few other studies available on the topic of how colours assigned by persons to the emotions are related to each other. In one of such studies [47], series of experiments were conducted with the synesthete who experiences emotion-evoked colours. During the experiments, an informal observation was made that the synesthete's colour for pride was a shade of blue and the colour for aggression was pinkish–red. Intriguingly, it was also noticed that his colour for arrogance was purple, presumably because the combination of blue and red in colour-space is purple, and the combination of pride and aggression in emotion–space is arrogance [47].

## 4 Conclusions

This article reported a case study that we conducted with two persons with colour synaesthesia. The main research goal was to establish how the colours chosen by the synesthetes for representing different emotions were related to each other. We decided to perform a case study with the synesthetes because according to the research literature (e.g., [23, 29-33]), the synesthetes have much stronger associations between emotions and colours compared with "ordinary" people. We make the following conclusions from the case study on choosing colours for emotions by the two synesthetes, which also answer the research questions:

- The colours chosen by the synesthetes for representing different emotions were idiosyncratic and culture-dependent. This finding indirectly supports the theory of constructed emotion [6-7], according to which emotions are social constructs that, as well as their colour mappings, are *supposed* to be constructed as different for different individuals and in different cultural environments.
- The colours that were chosen by the synesthetes for representing more complex emotions could be represented as combinations of the colours that the respective synesthete had chosen for representing the constituent emotions. This finding also indirectly supports the theory of constructed emotion [6-7], based on which more complex emotions are constructed from simpler emotions and affects [9-10], which was mirrored by colours chosen for the emotions by the synesthetes.
- The shades of the colours chosen by the synesthetes for representing emotions and even the choices of colours themselves depended on the context determined by the underlying narrative and its particular fragment.

The results of the case study can be used for employing colours for expressing emotions in protostories of IDNs and their variants. When doing so, it is important to design an IDN in such a manner that the starting point for using colours for expressing emotions could always be determined in a case-based way by assigning specific colours to the so-called basic emotions. An IDN should then be designed and programmed in such a way that colours standing for different emotions and their combinations could be adequately expressed in the course of the story. Preliminary ideas of how this could be done are presented in the article [48], which, however, deals with artificial sound-to-colour synaesthesia. The research work should be conducted in the direction of working out a kind of ontology [3] for emotions where the relationships between different emotions are semi-formally represented. This would yield a universal understanding of how affects [9-10] and simpler emotions form more complex emotions. That understanding, in turn, can be used for rendering even complex emotions with colours in interactive digital storytelling.

**Acknowledgements**

The authors express their gratitude for collaboration to the two synesthetes from Malaysia and Estonia, respectively, who participated in the case study reported in this paper.